\documentclass[twoside]{ilcws07}

\usepackage[latin1]{inputenc}

\usepackage[dvips]{graphicx,epsfig,color}

\usepackage{wrapfig,rotating}

\usepackage{amssymb,amsmath,array}

\usepackage{epsfig}

\begin{document}


\title{
Precise quark masses from sum rules} 



\author{Matthias Steinhauser
\vspace{.3cm}\\
Institut f{\"u}r Theoretische Teilchenphysik,
Universit{\"a}t Karlsruhe (TH)\\
76128 Karlsruhe, Germany
}



\maketitle

\begin{abstract}
In this contribution an improved analysis is described to extract
precise charm and bottom quark masses from experimental and theoretical
moments of the photon polarization function. The obtained
$\overline{\rm MS}$ mass values read $m_c(3~\mbox{GeV})=0.986(13)$~GeV and
$m_b(10~\mbox{GeV})=3.609(25)$~GeV.
\end{abstract}


\section{Introduction}

The theory of strong interaction has the strong coupling constant and the
quark masses as fundamental input parameters.
The latter constitute an essential input for the evaluation of weak decay
rates of heavy mesons and for quarkonium spectroscopy. Furthermore,
decay rates and branching ratios of a light Higgs boson --- as suggested by
electroweak precision measurements --- depend critically on the masses of
the charm and bottom quarks. Last not least, 
confronting the predictions for these masses with experiment is an 
important task for all variants of Grand Unified Theories.
To deduce the values in a consistent way from different experimental
investigations and with utmost precision is thus a must for current 
phenomenology.

The method described in this contribution goes back to
1977~\cite{Novikov:1977dq} and was applied to next-to-next-to-leading order
(NNLO) in Ref.~\cite{Kuhn:2001dm}. The NNNLO analysis, including updated
experimental input, was presented in Ref.~\cite{Kuhn:2007vp}.


\section{Moments}

The basic object which enters our analysis is the photon polarization function
defined through
\begin{eqnarray}
  \left(-q^2g_{\mu\nu}+q_\mu q_\nu\right)\,\Pi(q^2)
  &=&
  i\int {\rm d}x\,e^{iqx}\langle 0|Tj_\mu(x) j^{\dagger}_\nu(0)|0
  \rangle
  \,,
  \label{eq::pivadef}
\end{eqnarray}
with $j_\mu$ being the electromagnetic current. The normalized total cross
section for hadron production in $e^+e^-$ annihilation is then given by
\begin{eqnarray}
  R(s) &=& \frac{\sigma(e^+e^-\to\mbox{hadrons})}{\sigma_{\rm pt}}
  \,\,=\,\, 12\pi\,\mbox{Im}\left[ \Pi(q^2=s+i\epsilon) \right]
  \,,
\end{eqnarray}
where $\sigma_{\rm pt} = 4\pi\alpha^2/(3s)$. In the following we add a
subscript $Q$ to indicate the contribution from the heavy quark $Q$.

The idea for extracting a quark mass value $m_Q$ is based on moments constructed
from $\Pi_Q$. On one hand one can compute the Taylor expansion of $\Pi_Q(q^2)$
around $q^2=0$ and obtain the so-called ``theory-moments'' from
\begin{eqnarray}
  {\cal M}_n &=& \frac{12\pi^2}{n!}
  \left(\frac{{\rm d}}{{\rm d}q^2}\right)^n
  \Pi_Q(q^2)\Bigg|_{q^2=0}
  \,.
  \label{eq::mom-theory}
\end{eqnarray}
The three-loop contribution to $\Pi_Q(q^2)$ up to $n=8$ within QCD has been
computed in Refs.~\cite{Chetyrkin:1995ii,Chetyrkin:1997mb} and the four-loop
calculation for $n=0$ and $n=1$ has been performed in
Refs.~\cite{Chetyrkin:2006xg,Boughezal:2006px}. In the analysis of
Ref.~\cite{Kuhn:2007vp} also two-loop QED corrections and non-perturbative
contributions have been considered. The latter shows a visible effect only in
the case of the charm quark.

From dimensional considerations we have 
$m_Q\sim \left({\cal M}_n\right)^\frac{1}{2n}$ 
which implies a stronger dependence of $m_Q$ on variations of ${\cal M}_n$ for
smaller values of $n$. Furthermore, higher values of $n$ require a careful
theoretical treatment of the threshold region and the construction of an effective
theory. The analysis performed in Ref.~\cite{Kuhn:2007vp} is restricted to
$n=1,2,3$ and $4$. Note that precise mass values can only be obtained for the
three lowest moments since the non-perturbative contributions become too big
already for $n=4$.

One of the major advantages of the method discussed in this paper is that we can
adopt the $\overline{\rm MS}$ scheme for the quark mass entering
Eq.~(\ref{eq::mom-theory}) and thus directly extract the corresponding value
for the mass.

In order to extract experimental moments one exploits the analyticity of $\Pi_Q$
and arrives at
\begin{eqnarray}
  {\cal M}_n &=& \int \frac{{\rm d}s}{s^{n+1}} R_Q(s)
  \,,
  \label{eq::Mexp}
\end{eqnarray}
where $R_Q$ naturally divides into three parts:
  At lower energies one has the narrow resonances which are the $J/\Psi$
  and $\Psi^\prime$ for charm the $\Upsilon(nS)$ ($n=1,\ldots,4$) in the case
  of the bottom quark. The corresponding contributions to ${\cal M}_n$ are
  obtained with the help of the narrow width approximation for $R(s)$
  \begin{eqnarray}
    R^{\rm res}(s) &=& \frac{9\pi M_R \Gamma_{ee}}{\alpha^2}
    \left(\frac{\alpha}{\alpha(s)}\right)^2
    \delta(s-M_R^2)
    \,,
  \end{eqnarray}
  where the electronic widths $\Gamma_{ee}$ are known at the 1-2\% level.

  The second part is called threshold region and extends in the case of
  the charm quark from 3.73~GeV to about 5~GeV. In this region the cross
  section shows a rapid variation and can not be described by perturbation
  theory. Measurements from the BES collaboration from 2001~\cite{Bai:2001ct}
  and 2006~\cite{Ablikim:2006mb} provide excellent data for $R(s)$ with an
  uncertainty of about $4\%$. In order to obtain $R_c$ one has to subtract the
  contribution from the light quarks which is explained in detail in
  Ref.~\cite{Kuhn:2007vp}.

  The treatment of the bottom threshold region is quite similar.
  Measurements of $R$ from threshold up to 11.24~GeV have been
  performed by the CLEO Collaboration more than 20 years
  ago~\cite{Besson:1984bd}, with a systematic error of 6\%. 
  No radiative
  corrections were applied. The average value derived from the four data
  points below threshold amounts to $\bar R = 4.559 \pm 0.034 ({\rm stat.})$
  which is 28\% larger than the prediction from perturbative QCD (pQCD). However, a later
  result of CLEO~\cite{Ammar:1997sk} at practically the same energy,
  $R(10.52~\mbox{GeV})=3.56\pm0.01\pm0.07$, is
  significantly more precise and in perfect agreement with theory.
  Applying a rescaling factor of $1/1.28$ 
  to the old CLEO data not
  only enforces agreement between old and new CLEO data and pQCD 
  in the region below the $\Upsilon(4S)$, it leads, in addition, 
  also to excellent agreement between theory and experiment above 
  threshold around 11.2~GeV where pQCD
  should be applicable also to bottom production.
  Further support to our approach is provided by the CLEO measurement
  of the cross section for bottom quark production at
  $\sqrt{s}=10.865$~GeV which is given by
  $\sigma_b(\sqrt{s}=10.865~\mbox{GeV})=0.301\pm0.002\pm0.039$~nb~\cite{Huang:2006em}.   
  The central value can be converted to $R_b(10.865~\mbox{GeV})=0.409$.
  On the other hand, if one extracts $R_b(10.865~\mbox{GeV})$ from the
  rescaled CLEO data from 1984~\cite{Besson:1984bd} one obtains
  $R_b(10.865~\mbox{GeV})=0.425$ which deviates by less than 4\% from
  the recent result~\cite{Huang:2006em}.

\begin{figure}[t]
  \begin{center}
    \begin{tabular}{cc}
      \leavevmode
      \epsfxsize=.45\textwidth
      \epsffile{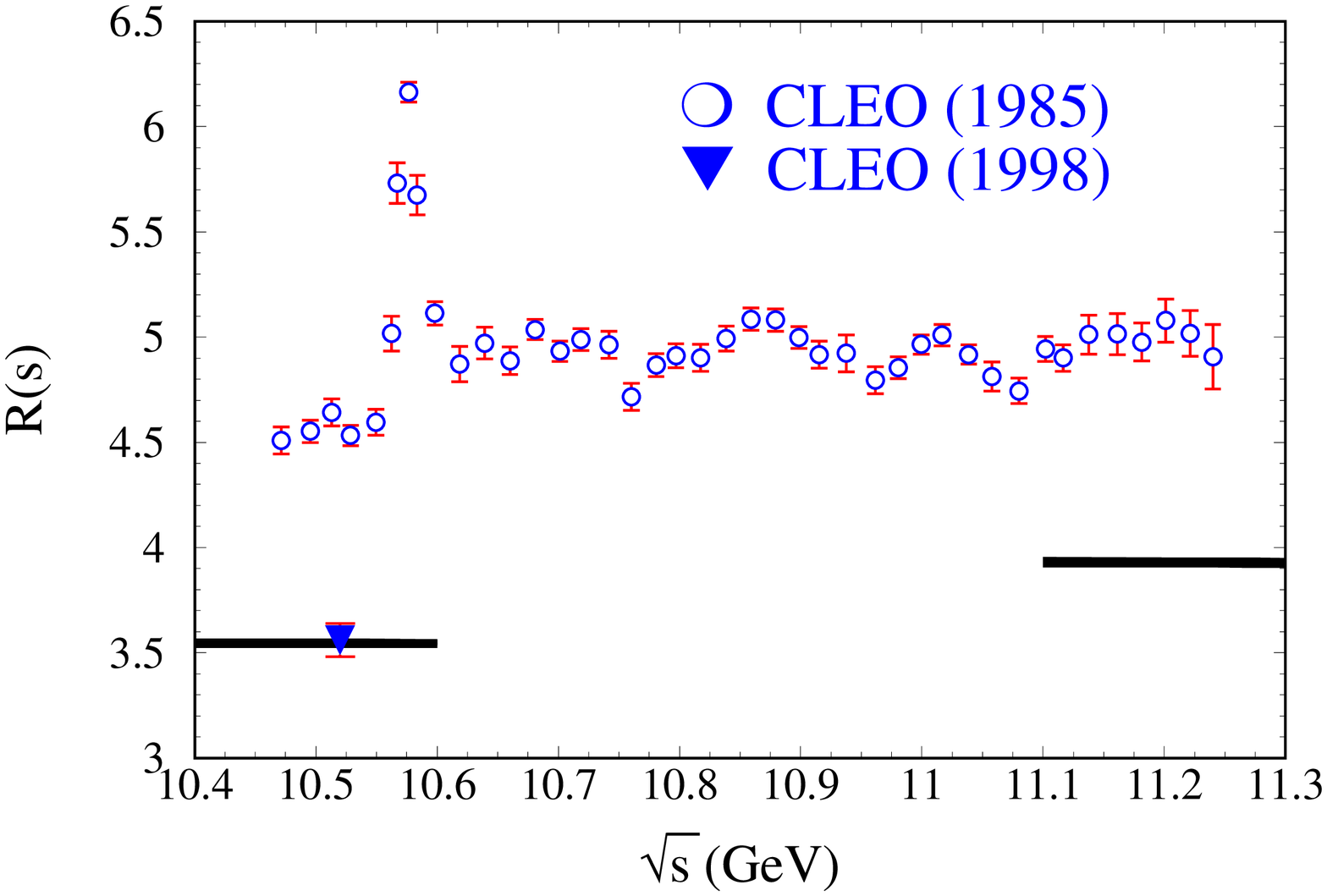}
      &
      \leavevmode
      \epsfxsize=.45\textwidth
      \epsffile{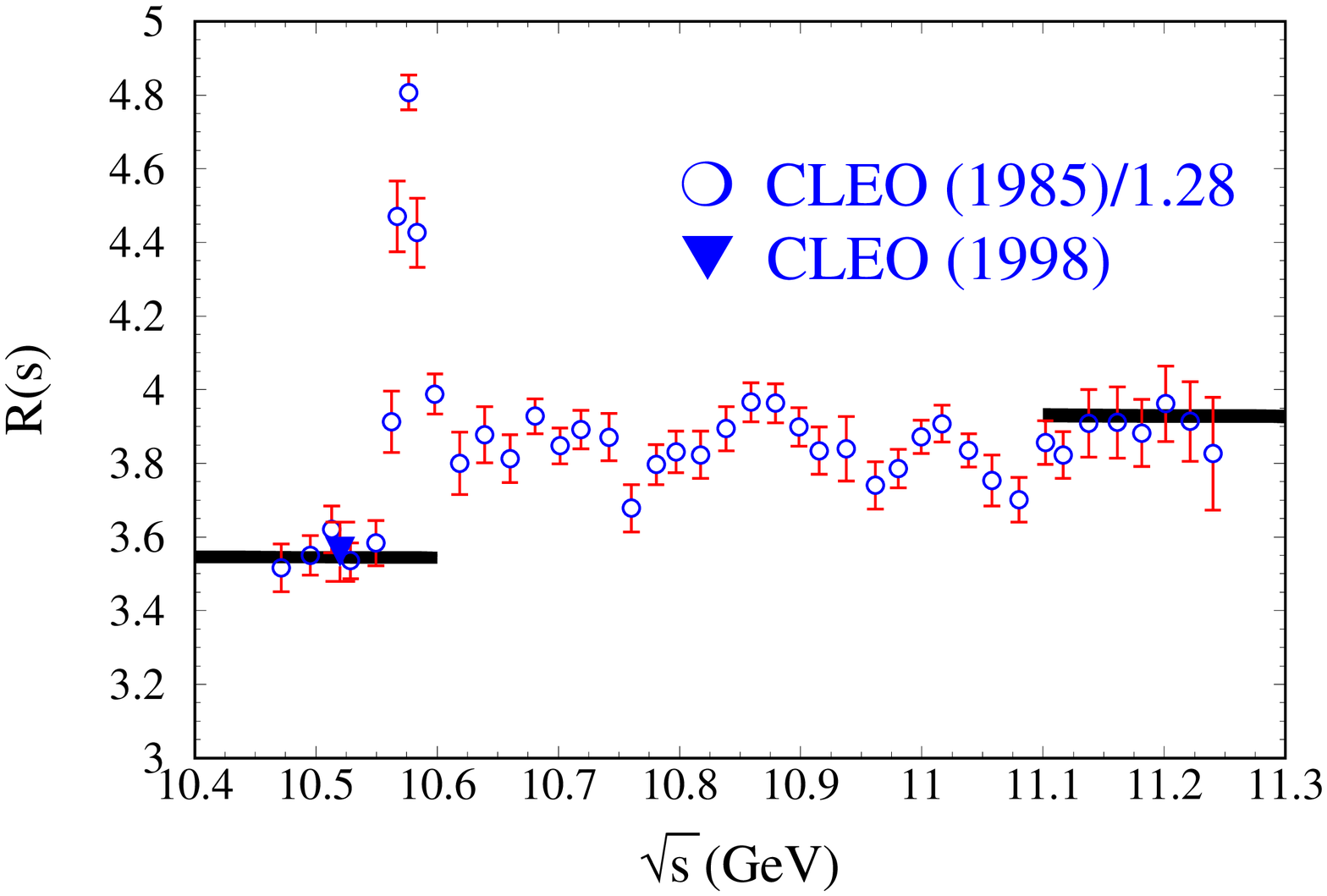}
    \end{tabular}
  \end{center}
  \vspace*{-2em}
  \caption{\label{fig::cleodata}
    In the left plot the data for $R(s)$ are shown as published in
    Refs.~\cite{Besson:1984bd} (circles) and~\cite{Ammar:1997sk}
    (triangle). The black curves are the predictions from pQCD outside
    the resonance region. In the right plot the older data
    from~\cite{Besson:1984bd} are rescaled by a factor 1/1.28.
          }
\end{figure}

  In Fig.~\ref{fig::cleodata} the original and the rescaled data
  from~\cite{Besson:1984bd} is shown and compared to pQCD and 
  data point from~\cite{Ammar:1997sk}.
  We thus extract the threshold contribution to the moments from the interval 
  10.62~GeV $\le \sqrt{s} \le$ 11.24~GeV 
  by applying the rescaling factor to the
  data, subtract the ``background'' from $u$, $d$, $s$ and $c$ quarks and
  attribute a systematic error of 10\% to the result.  

  The third contribution to the experimental moment is provided by the so-called
  continuum region which for the charm and bottom quark starts above 
  4.8~GeV and 11.24~GeV, respectively. In both cases there is no precise
  experimental data available. On the other hand, pQCD is supposed to work very
  well in these energy regions, in particular since $R_Q(s)$ is known to order
  $\alpha_s^2$ including the full quark mass dependence and to order
  $\alpha_s^3$ including quartic mass effects. For recent compilations we
  refer to Refs.~\cite{Chetyrkin:1996ia,Chetyrkin:1997pn,Harlander:2002ur}
  and would like to mention the {\tt Fortran} program {\tt
    rhad}~\cite{Harlander:2002ur} which provides a convenient platform to
  access easily the various radiative corrections.


\section{Quark masses}

\begin{figure}[t]
  \begin{center}
    \begin{tabular}{cc}
      \leavevmode
      \epsfxsize=0.45\textwidth
      \epsffile{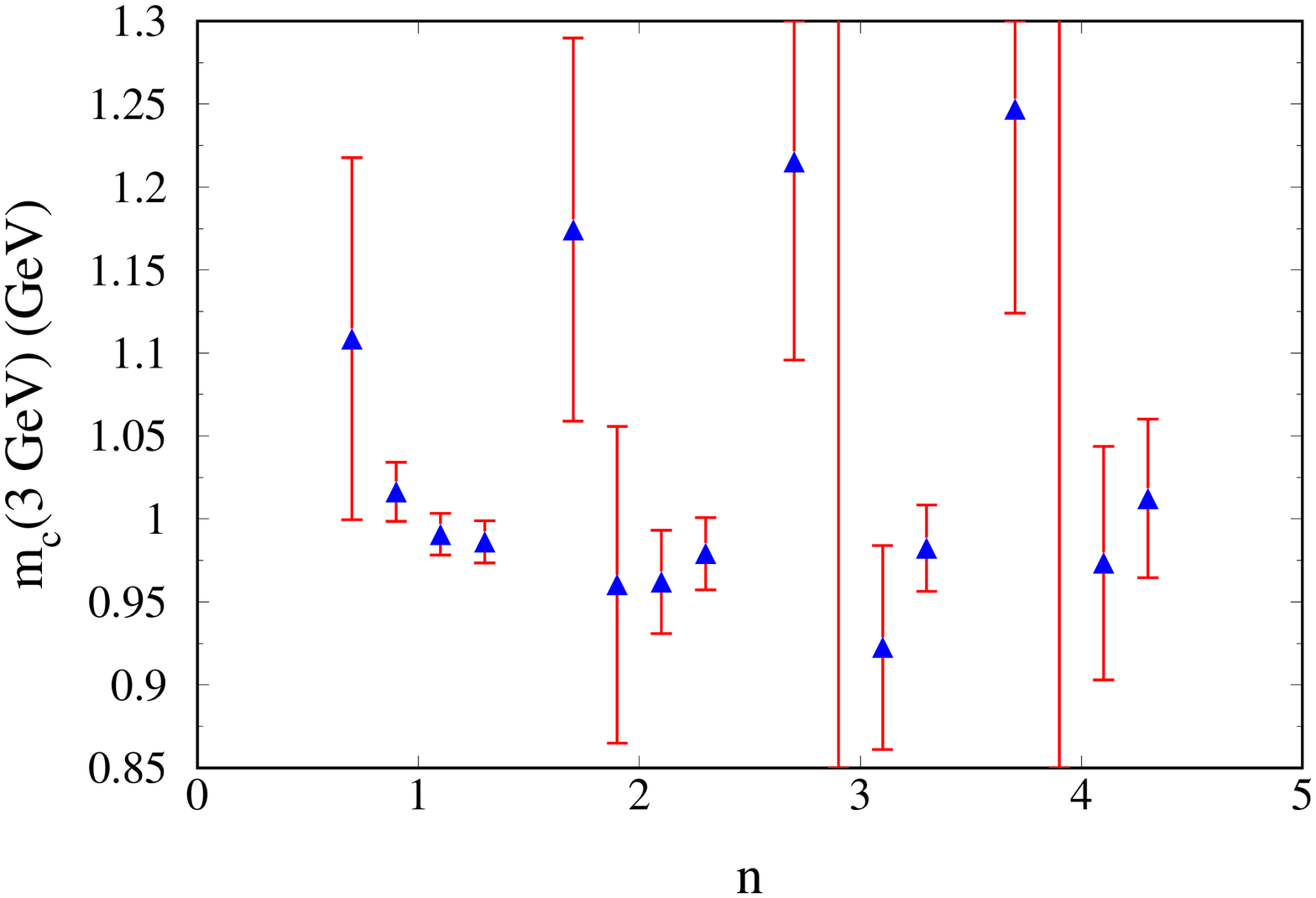}
      &
      \leavevmode
      \epsfxsize=0.45\textwidth
      \epsffile{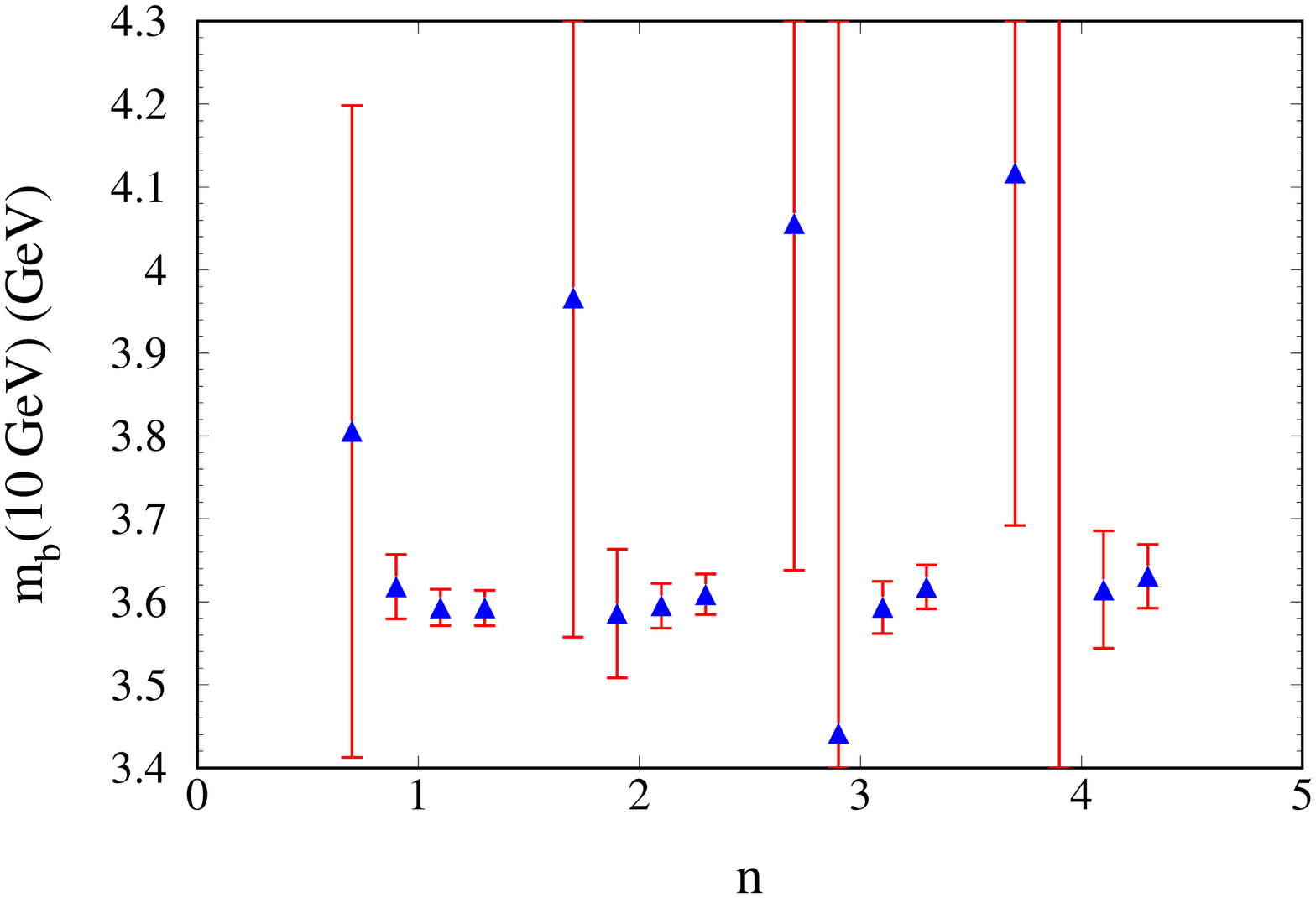}
    \end{tabular}
  \end{center}
  \vspace*{-2em}
  \caption{
    \label{fig::mom}$m_c(3~\mbox{GeV})$ (left) and $m_b(10~\mbox{GeV})$
    (right) for $n=1,2,3$ and $4$.
    For each value of $n$ the results from left to right correspond
    the inclusion of terms of order $\alpha_s^0$, $\alpha_s^1$,
    $\alpha_s^2$ and $\alpha_s^3$ to the theory-moments.
    Note, that for $n=3$ and $n=4$ the 
    uncertainties can not be determined
    in those cases where only the two-loop corrections of order $\alpha_s$ are
    included into the coefficients $\bar{C}_n$ as the equation cannot be
    solved for the quark mass.
  }
\end{figure}

Equating the theoretical and experimental moments
of Eqs.~(\ref{eq::mom-theory}) and~(\ref{eq::Mexp}), adopting $\mu=3$~GeV
($\mu=10$~GeV) for the charm (bottom) quark and solving for the quark
mass leads to the results which are shown in Fig.~\ref{fig::mom} in graphical
form.\footnote{The numerical results including a detailed error analysis can
  be found in Ref.~\cite{Kuhn:2007vp}.}
It is nicely seen that the results for $m_Q$ further stabilize when going from
three to four loops. At the same time the uncertainty is considerably reduced.
Furthermore, the preference for the first three moments is 
clearly visible.
Also the analysis for $n=2$ and $n=3$ leads to small errors, even if we
include the uncertainty from the yet uncalculated four-loop
contributions.\footnote{See Ref.~\cite{Kuhn:2007vp} for details.}
We emphasize the remarkable consistency between the three results which
we consider as additional confirmation of our approach.

The final result for the  $\overline{\rm MS}$-masses read
$m_c(3~\mbox{GeV})=0.986(13)$~GeV and $m_b(10~\mbox{GeV})=3.609(25)$~GeV. They
can be translated into $m_c(m_c)=1.286(13)$~GeV and $m_b(m_b)=4.164(25)$~GeV.
This analysis is consistent with but significantly more precise than a similar
previous study.

\section*{Acknowledgments}

I would like to thank Hans K\"uhn and Christian Strum for a fruitful
collaboration on this subject.
This work was supported the DFG through SFB/TR~9.


\begin{footnotesize}

\end{footnotesize}


\end{document}